\documentclass[prd, preprint, tightenlines, a4paper, eqsecnum, nofootinbib]{revtex4-1}
\usepackage{graphicx,color}
\usepackage{latexsym}
\usepackage{amsmath,amsfonts,amssymb}
\usepackage{physics}     
\usepackage{hyperref} 

\begin{document}


\title{Ground states of a Klein-Gordon field with \\ Robin boundary conditions in global anti-de Sitter spacetime}


\author{Claudio Dappiaggi}
\email{claudio.dappiaggi@unipv.it}
\affiliation{Dipartimento di Fisica, Universit\`a degli Studi di Pavia, Via Bassi, 6, 27100 Pavia, Italy}
\affiliation{Istituto Nazionale di Fisica Nucleare -- Sezione di Pavia, Via Bassi, 6, 27100 Pavia, Italy}

\author{Hugo R. C. Ferreira}
\email{hugo.ferreira@pv.infn.it}
\affiliation{Istituto Nazionale di Fisica Nucleare -- Sezione di Pavia, Via Bassi, 6, 27100 Pavia, Italy}

\author{Alessio Marta}
\email{alessio.marta@studenti.unimi.it}
\affiliation{Dipartimento di Fisica, Universit\`a degli Studi di Milano, Via Celoria 16, 20133 Milano, Italy}


\date{\today}

\begin{abstract}
We consider a real, massive scalar field both on the $n$-dimensional anti--de Sitter (AdS$_n$) spacetime and on its universal cover CAdS$_n$. In the second scenario, we extend the recent analysis on PAdS$_n$, the Poincar\'e patch of AdS$_n$, first determining all admissible boundary conditions of Robin type that can be applied on the conformal boundary. Most notably, contrary to what happens on PAdS$_n$, no bound state mode solution occurs. Subsequently, we address the problem of constructing the two-point function for the ground state satisfying the admissible boundary conditions. All these states are locally of Hadamard form being obtained via a mode expansion which encompasses only the positive frequencies associated to the global timelike Killing field on CAdS$_n$. To conclude we investigate under which conditions any of the two-point correlation functions constructed on the universal cover defines a counterpart on AdS$_n$, still of Hadamard form. Since this spacetime is periodic in time, it turns out that this is possible only for Dirichlet boundary conditions, though for a countable set of masses of the underlying field, or for Neumann boundary conditions, though only for even dimensions and for one given value of the mass.
\end{abstract}

\maketitle

\section{Introduction}

Quantum field theory on curved backgrounds is a rapidly developing branch of theoretical physics especially within the algebraic approach \cite{Benini:2013fia, Brunetti:2015vmh}. In the past few years several specific models have been thoroughly analyzed and important structural aspects have been deeply understood, {\it e.g.}, perturbative interactions,  renormalization theory and local gauge invariance. 

Yet an implicit assumption in many works is that the underlying background is globally hyperbolic. Such requirement has far reaching consequences both from the geometric and from the analytic point of view. In the first case it ensures that the causal structure of the spacetime does not encompass pathologies, such as closed causal curves. In the second case  it entails that wave like operators, such as the Klein-Gordon, the Dirac or the Proca equation, can be solved by assigning suitably regular initial data. As an additional consequence, whenever one considers a free field theory, one can follow a well-established quantization scheme, yielding an algebra of observables which encodes structural properties such as dynamics, locality and causality. The only freedom left is the choice of a quantum state of Hadamard form, a widely accepted condition which entails several relevant physical properties. On the one hand the quantum fluctuations of all observables are finite, while, on the other hand, it guarantees the existence of a covariant notion of Wick polynomials out of which one can deal with interactions within a perturbation scheme, see {\it e.g.} \cite{Kay:1988mu, Khavkine:2014mta}.

Nonetheless, although based on strong physical motivations, the hypothesis that the underlying spacetime $M$ is globally hyperbolic does not allow to consider several interesting phenomena and scenarios, the prime example being field theoretic models built on anti-de Sitter spacetime. This is a maximally symmetric solution of vacuum Einstein's equations with negative cosmological constant which has been at the heart of the renown AdS/CFT correspondence, see for example the recent monograph \cite{Ammon:2015wua}.

From the point of view of the quantization of free field theories, dropping the assumption of $M$ being globally hyperbolic, entails that any wave-like partial differential equation does not have necessary a well-posed initial value problem. As a consequence one can guarantee neither the existence nor the uniqueness of fundamental solutions for the operators ruling the dynamics. In turn this entails that one has no natural building block out of which imposing the canonical commutation relations. The reasons for such failure are manifold, but one can recognize two main sources of the problem: the existence of a (conformal) boundary and of closed causal curves.

Focusing on the first problem, one can observe that, whenever one considers wave-like operators, solutions can be constructed supplementing the initial data with suitable boundary conditions. On the contrary, the presence of closed causal curves leads to a more subtle issue since they entail that initial data can be associated unambiguously to solutions only if these are periodic along the pathological curves. 

In order to address if it is possible to find a way to circumvent all these problems, the most natural testing ground is the $n$-dimensional anti-de Sitter spacetime AdS$_n$. As a manifold this is not globally hyperbolic since it possesses both a (conformal) boundary and a periodic time direction. 

Our goal is to consider a massive, real scalar field on AdS$_n$, proving under which conditions it is possible to address the question of the existence of a coherent, covariant quantization scheme. This is certainly not the first paper on the topic, the first investigation on the issue dating the late seventies \cite{Avis:1977yn}. 

In order to disentangle the above two problems, our first step consists of considering CAdS$_n$, the universal cover of anti-de Sitter spacetime, which is a manifold still possessing a conformal boundary, but no closed timelike curve. In this setting it is known that the Klein-Gordon equation leads to a well-defined initial value problem, though most of the literature assumes only Dirichlet boundary conditions. For a rather exhaustive survey of the known results and approaches as well as for a collection of references on this topic, we refer to the following thesis \cite{Kent:2013}.

In a recent paper by two of us \cite{Dappiaggi:2016fwc}, it has been shown that, if one considers only the Poincar\'e patch of AdS$_n$, it is possible to use a mode decomposition together with techniques proper of Sturm-Liouville problems, in order to prove that one can consider a whole one-parameter family of boundary conditions of Robin type, which include as special case both the Dirichlet and the Neumann ones. In this work it has been shown that, for each of these boundary conditions, the Klein-Gordon equation can be solved in terms of initial data and unique fundamental solutions do exist. Hence canonical commutation relations can be imposed coherently. Yet, it turns out that, for a wide range of boundary conditions, the underlying mode solutions do encompass bound states. While, from a classical perspective, this is not a problem, it has rather drastic consequences at a quantum level. As a matter of fact, since the Poincar\'e patch possesses a global timelike Killing field, in \cite{Dappiaggi:2016fwc}, it has been studied the existence for each boundary condition of Robin type of ground states associated with the Klein-Gordon equation. It turned out that, while they do not exist whenever bound state mode solutions occur, in all other cases they can be constructed explicitly in terms of their associated two-point correlation function. In addition they enjoy several notable physical properties, such as the Hadamard condition.

The techniques used in \cite{Dappiaggi:2016fwc} are rather flexible; they have been studied from a rigorous viewpoint in \cite{Dappiaggi:2017wvj} and applied also to the analysis of a Klein Gordon field in BTZ spacetime in \cite{Bussola:2017wki}. In this paper, first we also apply them to the study of a massive, real scalar field in the global chart of CAdS$_n$, in order to investigate if the results obtained in the Poincar\'e patch do extend globally. The outcome of our analysis is partly surprising. While, on the one hand, we prove that Robin boundary conditions can be imposed, it turns out that bound state mode solutions never occur. As a consequence, since CAdS$_n$ is a static spacetime, we are able to construct explicitly, for each Robin boundary condition, the two-point function of the ground state. In addition, since, in the underlying mode decomposition, we consider only positive frequencies with respect to the underlying global timelike Killing field, it turns out that the Hadamard condition is automatically fulfilled. 

At last we investigate whether any of the two-point functions constructed defines a counterpart in AdS$_n$. To this end we have to cope with the time coordinate, associated to the global timelike Killing field, being periodic. In this respect, already in \cite{Avis:1977yn}, it was observed that such geometric feature entails that, for consistency, also the underlying two-point function must be periodic. This occurs only if the mass of the field assumes certain special values which form a countable set. Our first goal is to test such statement for arbitrary boundary conditions and not just for the Dirichlet ones as in \cite{Avis:1977yn}. As a result, we prove that, in addition to the solutions found in \cite{Avis:1977yn} no periodic two-point function exists except for one special value of the mass provided that we consider Neumann boundary conditions and even spacetime dimensions. As such, we conclude that, while the presence of (conformal) boundaries does not hinder the existence of a well-defined, full-fledged, covariant quantization scheme, the occurrence of closed timelike curves leads to severe restrictions on the parameters of the matter fields.

The paper is organized as follows: In Section \ref{section2}, first we recollect some basic geometric aspects of the $n$-dimensional anti-de Sitter spacetime AdS$_n$ and of its universal cover CAdS$_n$. Subsequently we consider the Klein-Gordon equation on CAdS$_n$ and we use a mode decomposition to construct an explicit basis of solutions. In Section \ref{section3} we revisit the dynamics within the framework of Sturm-Liouville theory, studying the most general class of boundary conditions of Robin type, which can be considered. In Section \ref{sec:Ground State} we show that, for each of these boundary conditions, it is possible to associate explicitly the two-point function of a ground state, which enjoys in addition the Hadamard property. Subsequently we investigate under which conditions any of such two-point functions yields a well-defined counterpart on AdS$_n$. Eventually we draw our conclusions. In the Appendix we discuss some more technical aspects concerning the construction of the two-point functions and we show, in particular, that no bound state mode solution occurs.

\section{Scalar field in AdS spacetime}
\label{section2}

In this section we recollect some basic geometric fact about the $n$-dimensional anti-de Sitter spacetime and we study the dynamics thereon of massive, real scalar field.

\subsection{The Geometry of AdS$_n$}

We consider AdS$_n$, $n\geq 3$, the maximally symmetric solution of vacuum Einstein's equations with negative cosmological constant $\Lambda$ \cite{Hawking:1973uf}. Such spacetime can be realized in $\mathbb{R}^{n+1}$, endowed with Cartesian coordinates $X_i$, $i=0,...,n$ and with the line element $\dd s^2=-\dd X^2_0-\dd X^2_1+\sum\limits_{i=2}^{n}\dd X^2_i$, as the hyperboloid
$$-X^2_0-X^2_1+\sum\limits_{i=2}^n X^2_i=-\ell^2,$$
where $\ell$ is related to $\Lambda$ via $\Lambda=-\frac{n(n+1)}{\ell}$. Henceforth we set $\ell=1$. For our purposes, it is mostly convenient to realize AdS$_n$ in terms of a global chart which we report for completeness:
\begin{equation}\label{eq:global_chart}
\left\{\begin{array}{l}
X_0 =\cosh\rho\cos t\\
X_1 = \cosh\rho\sin t\\
X_i = \sinh\rho\, e_i(\theta,\varphi_1,...,\varphi_{n-3})
\end{array}
\right.,
\end{equation}
where $i$ runs from $2$ to $n$, $t\in(0,2\pi)$, $\rho\in(0,\infty)$, $\theta\in (0,2\pi)$, while $\varphi_j\in(0,\pi)$ for all $j=1,...,n-3$. Here $e_i\equiv e_i(\theta,\varphi_1,...,\varphi_{n-3}) $ parametrizes a point on the unit $(d-2)$-sphere in terms of angular coordinates. In this representation and adopting henceforth the symbol $\underline{\theta}$ to indicate collectively all angular coordinates, {\it i.e.} $\underline{\theta}\equiv(\theta,\varphi_1,...,\varphi_{n-3})$, the line element of AdS$_n$ reads
\begin{equation}\label{eq:metric}
\dd s^2=-\cosh^2\rho\,\dd t^2+\dd \rho^2+\sinh^2\rho\, \dd\mathbb{S}^2_{n-2}(\underline{\theta}),
\end{equation}
where $\dd \mathbb{S}^2_{n-2}$ stands for the standard line element of the unit $(n-2)$-sphere. Observe that, following \eqref{eq:global_chart}, the time direction is periodic and, for this reason, it is often convenient to consider the universal covering of anti-de Sitter spacetime, which we refer as CAdS$_n$ and whose line element is nothing but \eqref{eq:metric}, though with $t\in\mathbb{R}$. As a last remark, we recall that both AdS$_n$ and CAdS$_n$ possess a conformal, timelike, boundary which, in the chosen chart, can be heuristically built by considering $\rho\to\infty$. More precisely, starting from \eqref{eq:metric} and multiplying the metric by the conformal factor $\Omega^2=\frac{1}{\cosh^2\rho}$, via the coordinate transformation $\rho\to r\equiv r(\rho)$ defined out of $\cosh\rho=\frac{1}{\cos r}$, the conformally rescaled line element becomes 
$$\Omega^2 \dd s^2=-\dd t^2+\dd r^2+\sin^2 r\,\dd\mathbb{S}^2_{n-2}(\underline{\theta}).$$
Since $r\in (0,\frac{\pi}{2})$ we have realized the universal covering of anti-de Sitter spacetime as an open subset of the $n$-dimensional Einstein static Universe and we can thus attach a conformal boundary to CAdS$_n$ as $r=\frac{\pi}{2}$. The same holds true for AdS$_n$ though keeping the time coordinate $t$ periodic.

\subsection{Klein-Gordon equation}

Although our ultimate goal is the construction of the two-point function of the ground state of a massive, real scalar field on AdS$_n$, it is more convenient to work directly on CAdS$_n$, unless state otherwise. Hence, let us consider $\Phi:\textrm{CAdS}_n\to\mathbb{R}$ which satisfies the Klein-Gordon equation
\begin{subequations}
\begin{gather}	
P\Phi=\left(\Box_g-m^2_0-\xi R\right)\Phi=0,\label{eq:KG}\\
\Box_g=-\frac{\partial^2_t}{\cosh^2\rho}+\partial^2_\rho+F_{n-2}(\rho)\partial_\rho+\frac{\Delta_{\mathbb{S}_{n-2}}}{\sinh^2\rho}\label{eq:KG2}
\end{gather}
\end{subequations}
where $\Box_g$ is the D'Alembert wave operator built out of \eqref{eq:metric}, $\Delta_{\mathbb{S}_{n-2}}$ is the Laplacian on the unit $(n-2)$-sphere, $F_{n-2}(\rho)=\partial_\rho\ln(\cosh\rho\sinh^{n-2}\rho)$. Furthermore $m^2_0\geq 0$,  $R=-n(n-1)$ is the scalar curvature and $\xi\in\mathbb{R}$. Equation \eqref{eq:KG2} has been thoroughly studied by several authors starting from the first investigation in a four dimensional scenario in \cite{Avis:1977yn}, later extended in \cite{Burgess:1984ti} and in  \cite{Cotabreveescu:1999em,Ambrus:2018olh}.  To construct the solutions of \eqref{eq:KG2}, it is convenient to replace the coordinate $\rho$ with $z\doteq 1-\frac{1}{\cosh^2\rho}\in(0,1)$. In addition the field admits the expansion
$$\Phi(t,z,\underline{\theta})=\sum_{\underline{L}}\int\limits_{\mathbb{R}}d\omega\;\Phi_{\omega,\underline{L}}(z)Y_{\underline{L}}(\underline{\theta})e^{-i\omega t},$$
where $Y_{\underline{L}}(\underline{\theta})=Y_{l_1,...,l_{n-2}}(\underline{\theta})$ are the real scalar spherical harmonics on $\mathbb{S}_{n-2}$, {\it cf.} \cite{Higuchi:1986wu}, with $l_{n-2}\geq l_{n-3}\geq...\geq l_2\geq|l_1|$ and such that $\Delta_{\mathbb{S}_{n-2}}Y_{\underline{L}}(\underline{\theta})=l_{n-2}(l_{n-2}+n-3)Y_{\underline{L}}(\underline{\theta})$. Consequently $\sum_{\underline{L}}$ is a short cut for $\sum\limits_{l_{n-2}=1}^\infty\sum\limits_{l_{n-3}=1}^{l_{n-2}}...\sum\limits_{l_1=-l_2}^{l_2}$. In the special case $n=3$, observe that we are left with the Fourier series with respect to the sole angular coordinate $\theta$ and $l\equiv l_1\in\mathbb{Z}$. Therefore \eqref{eq:KG} reduces to the following ODE
\begin{gather}
K\Phi_\omega(z)=0,\notag\\
K=4z(1-z)\frac{d^2}{dz^2}+Q_1(z)\frac{d}{dz}
-Q_2(z)+\omega^2,\label{eq:ODE}
\end{gather}
where
\begin{subequations}
	\begin{align}
	Q_1(z)=2(n-1)-4z,\\
	Q_2(z)=\frac{M^2}{1-z}+\frac{l_{n-2}(l_{n-2}+n-3)}{z},
	\end{align}
\end{subequations}
where $M^2=m^2_0+\xi R$. Observe that, for later convenience, we will henceforth make explicit only the dependence on $\omega$ of all functions. Using Frobenius method to study the asymptotic behaviour of the solutions near the end points $z=0$ and $z=1$ suggests to make the ansatz
$$\Phi_\omega(z)=z^{\alpha_+}(1-z)^{\beta_+}f_\omega(z),$$
where we consider the positive roots of $4\alpha^2-2\alpha(3-n)-l_{n-2}(l_{n-2}+n-3)=0$ and $4\beta^2-2(n-1)\beta-M^2=0$, that is
\begin{subequations}
\begin{align}
\alpha_+=\frac{l_{n-2}}{2},\label{eq:alpha+}\\
\beta_+=\frac{1}{4}(n-1 + 2\nu),\label{eq:beta+}
\end{align}
\end{subequations}
where, for later notational convenience, we introduced the parameter
\begin{equation}\label{eq:nu}
\nu\doteq\frac{1}{2}\sqrt{(n-1)^2 + 4 M^2},
\end{equation}
In the special case $n=3$, observe that, since $l_1\in\mathbb{Z}$, we set $2\alpha_+=|l_1|$. Observe that, for \eqref{eq:beta+} to be well-defined, we need to require that $4 M^2 + (n-1)^2\geq 0$. The lowest admissible value for $M^2$ corresponds to the Breitenlohner-Freedman (BF) bound \cite{Breitenlohner:1982jf}. The extremal case $4M^2=-(n-1)^2$ has a special behaviour and it should be analysed on its own. In this paper we will not consider further this case. The remaining unknown $f_{\omega,l_{n-2}}(z)$ satisfies the hypergeometric differential equation
$$z(1-z)f^{\prime\prime}_\omega+(c-(a+b+1)z)f^\prime_\omega-abf_\omega=0,$$
where the prime symbol stands for the derivative with respect to $z$, while
\begin{subequations}
	\begin{gather}
	a=\alpha_++\beta_+-\frac{\omega}{2},\label{eq:a}\\
	b=\alpha_++\beta_++\frac{\omega}{2},\label{eq:b}\\
	c=l_{n-2}+\frac{n-1}{2}.\label{eq:c}
	\end{gather}
\end{subequations}
Depending on the end-point that one wishes to investigate, different basis of solutions of \eqref{eq:ODE} are convenient. In the case of $z=0$ we consider 
\begin{widetext}
\begin{gather}
\Phi_{1,\omega}(z)=z^{\frac{l_{n-2}}{2}}(1-z)^{\beta_+}F_2^1(a,b,c;z)\label{Phi1}\\
\Phi_{2,\omega}(z)=z^{\frac{3-n-l_{n-2}}{2}}(1-z)^{\beta_+}F_2^1(a-c+1,b-c+1,2-c;z)\label{Phi2}
\end{gather}
\end{widetext}
where $\Phi_{2,\omega}$ is linearly independent from $\Phi_{1,\omega}$ provided that $c\notin\mathbb{N}$ which occurs only for even spacetime dimensions. If $n$ is odd, then \eqref{Phi2} must be replaced with a different function, whose form depends whether $a$ is a positive integer or not. As we will discuss in the next section, these solutions will play no role in our investigation. Hence, we shall not write them explicitly, although an interested reader can find them in \cite[\S 15.10]{NIST}. Observe that, if $n=3$ then, $l_1$ should be replaced with $|l_1|$.

On the contrary if $z=1$, we consider the following basis of solutions of \eqref{eq:ODE}
\begin{widetext}
\begin{gather}
\Phi_{3,\omega}(z)=z^{\alpha_+}(1-z)^{\beta_+}F_2^1(a,b,a+b+1-c;1-z),\label{Phi3}\\
\Phi_{4,\omega}(z)=z^{\alpha_+}(1-z)^{-\beta_++\frac{n-1}{2}}F_2^1(c-a,c-b,c-a-b+1;1-z),\label{Phi4}
\end{gather}
\end{widetext}
which is admissible provided that $a+b+c-1$ is not an integer. In this case $\Phi_4(z)$ must be replaced with another linearly independent solution whose explicit form is listed in \cite[\S 15.10]{NIST}. As in the previous case, these exceptions will play no role in the following discussion and hence we avoid reporting them explicitly.

\section{Boundary conditions}
\label{section3}

Having established a basis of the solutions of \eqref{eq:ODE} both at $z=0$ and at $z=1$, we can ask ourselves if and which boundary conditions should be imposed at both ends. To answer this question we follow the same procedure as in \cite{Dappiaggi:2016fwc,Dappiaggi:2017wvj,Bussola:2017wki} which relies on Sturm-Liouville theory for ordinary differential equations. A reader interested in more details can consult \cite{Zettl:2005} on which we base our analysis. The first step calls for rewriting \eqref{eq:ODE} in an equivalent Sturm-Liouville form, namely
\begin{gather}
	S\Phi_\omega=0\notag\\
	S=\frac{\dd}{\dd z}\left(P(z)\frac{\dd}{\dd z}\right)+\widetilde{Q}(z)-\omega^2\mathcal{J}(z),\label{eq:SL}
\end{gather} 
where $P(z)=-Q_1(z)\mathcal{J}(z)$, $\widetilde{Q}(z)=Q_2(z)\mathcal{J}(z)$ and
\begin{equation}\label{eq:measure}
\mathcal{J}(z)\doteq\frac{z^{\frac{n-3}{2}}}{2(1-z)^{\frac{n+1}{2}}}.
\end{equation}

The second step consists of establishing under which constraints \eqref{Phi1} and $\eqref{Phi2}$ lie in $L^2((0,z_0);\dd\mu(z))$ while \eqref{Phi3} and \eqref{Phi4} lie $L^2((z^\prime_0,1),\dd\mu(z))$, $z_0,z^\prime_0$ being two arbitrary points in $(0,1)$ while $\dd\mu(z)=\mathcal{J}(z)\dd z$.

Starting from $z=0$ a direct inspection of \eqref{Phi1} and \eqref{Phi2} unveils that their asymptotic behaviour is respectively dominated by $z^{\alpha_+}$ and $z^{\frac{3-n-l_{n-2}}{2}}$. Taking into account \eqref{eq:measure}, it descends that 
$$\Phi_{1,\omega}(z)\in L^2((0,z_0);\dd\mu(z))\Longleftrightarrow l_{n-2}+\frac{n-3}{2}>-1,$$
which is always true. On the contrary
$$\Phi_{2,\omega}(z)\in L^2((0,z_0);\dd\mu(z))\Longleftrightarrow -l_{n-2}-\frac{n-3}{2}>-1,$$
which is never valid unless $n=3,4$ and $l_{n-2}=0$. Hence, since we do not want to admit different boundary conditions for different values of $l_{n-2}$, at $z=0$ only \eqref{Phi1} is admissible. Observe that this statement justifies our claim in the previous section that there is no need to study in detail the alternative expressions of \eqref{Phi2} which occur when $c$ as in \eqref{eq:c} in integer valued.

Let us now focus on $z=1$. In this case a direct inspection of \eqref{Phi3} and \eqref{Phi4} show that the asymptotic behaviour of the solutions is dominated respectively by $(1-z)^{\beta_+}$ and by $(1-z)^{\beta_++\frac{n-1}{2}}$. Taking also into account both \eqref{eq:measure} and \eqref{eq:beta+}, it holds that
$$\Phi_{3,\omega}(z)\in L^2((z^\prime_0,1);\dd\mu(z))\Longleftrightarrow \nu>-1,$$
where $\nu$ is defined in \eqref{eq:nu}. The inequality is always fulfilled due to the Breitenlohner-Freedman bound. Hence \eqref{Phi3} is always admissible and, following the nomenclature proper of Sturm-Liouville theory, we shall call it {\em principal solution}, since it tends to $0$ as $z\to 1$ faster than any other solution of \eqref{eq:ODE} which is not a scalar multiple of $\Phi_{3,\omega}(z)$. At the same time, still taking into account both \eqref{eq:measure} and \eqref{eq:beta+},
$$\Phi_{4,\omega}(z)\in L^2((z^\prime_0,1);\dd\mu(z))\Longleftrightarrow \nu<1,$$

We observe that, whenever $0<\nu<1$ the quantity $c-a-b+1$ cannot be integer valued. This justifies our claim in the previous section that there is no need to consider the alternative forms of \eqref{Phi4}. In view of our results and using still the nomenclature proper of Sturm-Liouville theory, we call $z=0$ {\em limit point} and no boundary condition should be assigned there being only \eqref{Phi1} admissible. On the contrary
\begin{enumerate}
	\item if $\nu\geq 1$ only the principal solution \eqref{Phi3} is admissible at $z=1$. Hence no boundary condition is necessary and also $z=1$ is limit point.
	\item if $0<\nu<1$ then both \eqref{Phi3} and \eqref{Phi4} are admissible. In this case $z=1$ is called {\em limit circle} and it is necessary to impose a boundary condition. More precisely we say that $\Phi_{\gamma,\omega}$, solution of \eqref{eq:ODE}, satisfies a Robin boundary condition, if there exists $\gamma\in [0,\pi)$ such that 
	\begin{equation}\label{eq:boundary_condition}
	\lim\limits_{z\to 1}\left(\cos\gamma\, W_z[\Phi_{\gamma,\omega},\Phi_{3,\omega}]+\sin\gamma\, W_z[\Phi_{\gamma,\omega},\Phi_{4,\omega}]\right)=0,
	\end{equation}
	where $W_z[\Phi_{\gamma,\omega},\Phi_{i,\omega}]\doteq\frac{d\Phi_{\gamma,\omega}}{dz}\Phi_{i,\omega}-\Phi_{\gamma,\omega}\frac{d\Phi_{i,\omega}}{dz}$, $i=3,4$, is the Wronskian between $\Phi_{\gamma,\omega}$ and $\Phi_{i,\omega}$. Hence, up to a multiplicative and irrelevant constant we can set up
	\begin{equation}\label{Phi_gamma}
	\Phi_{\gamma,\omega}(z)=\cos\gamma\,\Phi_{3,\omega}(z)+\sin\gamma\,\Phi_{4,\omega}(z),
	\end{equation}
	where $\Phi_{3,\omega}$ and $\Phi_{4,\omega}$ are taken as in \eqref{Phi3} and \eqref{Phi4} respectively.
\end{enumerate}

 Recalling that \eqref{Phi3} is the principal solution, this justifies that we refer to the case $\gamma=0$ as {\em Dirichlet boundary condition}, while to that for which $\gamma=\frac{\pi}{2}$ as {\em Neumann boundary condition}. Observe that, while the former relies on the unambiguous choice of the principal solution, the latter is based on selecting any other solution of \eqref{eq:ODE} which is both square-integrable and linearly independent from \eqref{Phi3}. For this reason the Neumann boundary condition is not a universal concept contrary to the Dirichlet counterpart.

\section{Ground State}
\label{sec:Ground State}

\subsection{Two Point Function in CAdS$_n$}
\label{sec:tpfCAdS}

In this section we discuss the existence of a ground state for a massive, real scalar field obeying \eqref{eq:KG} on CAdS$_n$ for each admissible boundary conditions of Robin type classified in the previous section. To start with we will only consider the universal cover of the $n$-dimensional anti-de Sitter spacetime, in order to avoid any issue with the time coordinate being periodic. The construction of a ground state has been already discussed in the literature by several research groups, though only the Dirichlet boundary condition has been considered. Different construction methods have been outlined in \cite{Allen:1985wd,Camporesi:1991nw,Kent:2014nya}, though we shall be employing a mode expansion, which has been first considered in \cite{Burges:1985qq} with Dirichlet boundary conditions. The following discussion complements that in \cite{Dappiaggi:2016fwc} where the ground state for a massive real scalar field with arbitrary boundary conditions of Robin type has been constructed in the Poincar\'e patch of an $n$-dimensional anti-de Sitter spacetime.

In the following, by {\em two-point function} (or Wightman function) we refer to a bidistribution $\lambda_2\in\mathcal{D}^\prime(\textrm{CAdS}_n\times\textrm{CAdS}_n)$ such that
\begin{equation}\label{eq:eom_2pt}
(P\otimes\mathbb{I})\lambda_2=(\mathbb{I}\otimes P)\lambda_2=0 \, ,
\end{equation}
where $P$ is defined in \eqref{eq:KG} and
\begin{equation}\label{eq:pos_2pt}
\lambda_2(f,f)\geq 0 \, , \quad \forall f\in C^\infty_0(\textrm{CAdS}_n) \, .
\end{equation}
In addition, the antisymmetric part of $\lambda_2$ is constrained to coincide with the commutator distribution, in order to account for the canonical commutation relations (CCRs) of the underlying quantum field theory.

In order to make this last requirement explicit, let us consider the coordinate system $(t,z,\underline{\theta})$ introduced in \eqref{eq:metric} with $\rho$ replaced by $z$. Working at the level of the integral kernel for $\lambda_2$, imposing the CCRs is tantamount to requiring that the antisymmetric part  $iG(x,x^\prime)$, $x,x^\prime\in\textrm{CAdS}_n$, where 
$$ i G(x,x^\prime) = \lambda_2(x,x^\prime)-\lambda_2(x^\prime,x)$$
satisfies \eqref{eq:eom_2pt} together with the initial conditions
\begin{subequations} \label{eq:initial_conditions_E}
	\begin{align} 
	G(x,x^\prime)|_{t=t^\prime} &= 0, \label{eq:initial_conditions_E_1} \\
	\partial_t  G(x,x^\prime)|_{t=t^\prime} &=-\partial_{t^\prime}G(x,x^\prime)|_{t=t^\prime}=\frac{\delta(z-z^\prime)\delta(\underline{\theta}-\underline{\theta}^\prime)}{\mathcal{J}(z)}, \label{eq:initial_conditions_E_2}
	\end{align}
\end{subequations}
with $\mathcal{J}(z)$ as in \eqref{eq:measure}. Here $\delta(\underline{\theta}-\underline{\theta}^\prime)$ is a compressed form for $\delta(\theta-\theta^\prime)\prod_{i=1}^{n-3}\delta(\varphi_i-\varphi^\prime_i)$. 
In order to build explicitly \eqref{eq:pos_2pt}, it suffices to focus on the case $\nu\in (0,1)$, $\nu$ being defined in \eqref{eq:nu}. In this case, \eqref{eq:boundary_condition} entails that we can consider a one-parameter family of boundary conditions ruled by $\gamma\in [0,\pi)$, to each of which it corresponds a different two-point correlation function. Most notably, if we set $\gamma=0$, our analysis applies to all values of $\nu$, including the regime $\nu\geq 1$, which, therefore, we do not need to discuss in detail.

In view of the invariance of the metric under rotations and time translations, we can make the following ansatz for the integral kernel of $\lambda_2$:
\begin{widetext}
\begin{equation}\label{eq:integral_kernel}
\lambda_2(x,x^\prime)=\lim\limits_{\epsilon\to 0^+}\sum_{\underline{L}}\int\limits_0^\infty d\omega\,e^{-i\omega(t-t^\prime-i\epsilon)}Y_L(\underline{\theta})Y_L(\underline{\theta}^\prime)\widehat{\lambda}_{2,L,\omega}(z,z^\prime),
\end{equation}
\end{widetext}
where $i\epsilon$ is a suitable regularization and the limit has to be taken in the weak sense. Recall that $Y_L(\underline{\theta})$ are the real scalar spherical harmonics on the $(n-2)$-sphere. In \eqref{eq:integral_kernel} we have considered only positive frequencies since we aim at constructing the two-point function of a ground state. A direct comparison between \eqref{eq:integral_kernel} and both \eqref{eq:initial_conditions_E_1} and \eqref{eq:initial_conditions_E_2} unveils that the initial conditions for the antisymmetric part of $\lambda_2$ are automatically satisfied if 
\begin{equation}\label{eq:delta_form}
\int\limits_{\mathbb{R}}d\omega\,\omega\widehat{\lambda}_{2,L,\omega}(z,z^\prime)=\frac{\delta(z-z^\prime)}{\mathcal{J}(z)},
\end{equation}
where $\mathcal{J}(z)$ is defined in \eqref{eq:measure}. In addition \eqref{eq:eom_2pt} entails that 
$$(S\otimes\mathbb{I})\widehat{\lambda}_{2,L,\omega}=(\mathbb{I}\otimes S)\widehat{\lambda}_{2,L,\omega}=0,$$
where $S$ is the Sturm-Liouville form \eqref{eq:SL} of \eqref{eq:ODE}. Using this last equation and \eqref{eq:delta_form}, we can employ the spectral calculus for $S$ in order to derive an explicit form for $\widehat{\lambda}_{2,L,\omega}$ in terms of the solutions of \eqref{eq:ODE}. Since this is a lengthy and technical calculation we postpone it to the Appendix, so not to disrupt the flow of this section. Hence, using \eqref{eq:Delta_resolution}, it holds that, whenever $\nu\in(0,1)$
\begin{widetext}
\begin{eqnarray}
\lambda_{2,\gamma}(x,x^\prime)
=\lim\limits_{\epsilon\to 0^+}\sum_{k=0}^\infty\sum_{\underline{L}}\,e^{-i\omega_{k,\gamma,+}(t-t^\prime-i\epsilon)}\left(\cos\gamma \ C(\omega_{k,\gamma,+})+\sin\gamma \ D(\omega_{k,\gamma,+})\right)\notag\\
\times\Phi_{1,\omega_{k,\gamma,+}}(z)\Phi_{1,\omega_{k,\gamma,+}}(z^\prime)Y_L(\underline{\theta})Y_L(\underline{\theta^\prime}),\label{eq:2pt_R}
\end{eqnarray}
\end{widetext}
where $\gamma\in(0,\pi)$, $\gamma\neq\frac{\pi}{2}$, in the second line all quantities which are implicitly dependent on the frequency are evaluated for $\omega=\omega_{k,\gamma,+}$. To conclude we need to write also the integral kernel of the two-point function in the case of Dirichlet and Nuemann boundary conditions. Using \eqref{eq:Delta_resolution_D} and \eqref{eq:Delta_resolution_N} respectively one obtains
\begin{widetext}
	\begin{gather}\label{eq:2pt_D}
	\lambda_{2,0}(x,x^\prime)=\lim\limits_{\epsilon\to 0^+}\sum_{k=0}^\infty\sum_{\underline{L}}\,e^{-i\omega_{k,0,+}(t-t^\prime-i\epsilon)}C_0(\omega_{k,0,+})\Phi_{1,\omega_{k,0,+}}(z)\Phi_{1,\omega_{k,0,+}}(z^\prime)Y_L(\underline{\theta})Y_L(\underline{\theta}^\prime),\\
	\lambda_{2,\frac{\pi}{2}}(x,x^\prime)
	=\lim\limits_{\epsilon\to 0^+}\sum_{k=0}^\infty\sum_{\underline{L}}\,e^{-i\omega_{k,\frac{\pi}{2},+}(t-t^\prime-i\epsilon)}D_{\frac{\pi}{2}}(\omega_{k,\frac{\pi}{2},+})\Phi_{1,\omega_{k,\frac{\pi}{2},+}}(z)\Phi_{1,\omega_{k,\frac{\pi}{2},+}}(z^\prime)Y_L(\underline{\theta})Y_L(\underline{\theta}^\prime),\label{eq:2pt_N}
	\end{gather}
\end{widetext}
where $\omega_{k,0,+}$ are listed in \eqref{eq:D_frequencies} while $\omega_{k,\frac{\pi}{2},+}$ in \eqref{eq:N_frequencies}. Observe that, if we consider the regime $\nu\geq 1$, then no boundary condition is necessary and the only ensuing two-point function has the form of \eqref{eq:2pt_D}. In addition we remark two notable properties of $\lambda_{2,\gamma}$ with $\gamma\in [0,\pi)$. On the one hand, since all these two-point correlation functions are built out of positive frequencies with respect to a global timelike Killing field, they are ground states, hence of Hadamard form as proven in full generality in \cite{Sahlmann:2000fh}. For this reason each $\lambda_{2,\gamma}$ is a legitimate starting point to construct Wick ordered observables, such as in particular the regularized stress-energy tensor. On the other hand, in comparison to their counterpart on the Poincar\'e patch built in \cite{Dappiaggi:2016fwc}, it turns out that no bound state mode solution occurs. Most notably, it turns out that, restricting the attention to the Poincar\'e patch of an $n$-dimensional anti-de Sitter spacetime, the same problem considered in this paper leads to discovering that, for half of the boundary conditions of Robin type, bound state mode solutions occur, corresponding to purely imaginary frequencies in the resolution of the Dirac delta distribution. Hence no ground state exists for these particular scenarios. In the context considered in this paper, since all admissible frequencies which occur are real, such pathological feature apparently does not exist.

\subsection{Two Point Function in AdS$_n$: Mass Constraints}
\label{sec:eigenfreq}

In the previous section we have constructed the integral kernel of the two-point function of the ground state for a massive real scalar field in CAdS$_n$ with arbitrary boundary conditions of Robin type. We can now investigate if any of these correlation functions defines a counterpart on AdS$_n$. In this case we have to account for the time coordinate $t$ being periodic of period $2\pi$, see \eqref{eq:metric}. 

In order for \eqref{eq:integral_kernel} to be compatible with this geometric constraint, it is necessary to start from \eqref{eq:integral_kernel}, constructing a counterpart periodic in the variable $t$. Yet this procedure has the net disadvantage that, being all admissible two-point correlation functions singular, making them periodic would create in general a bidistribution with additional singularities, not compatible with the Hadamard condition. The only possible exception to this pathological scenario occurs if the frequencies in the mode expansion of \eqref{eq:integral_kernel} are integer valued. A direct investigation of the two-point functions for all $\gamma\in[0,\pi)$ unveils the following constraints on the admissible values for the masses of the Klein-Gordon field:

\begin{enumerate}
\item Imposing Dirichlet boundary condition, \eqref{eq:D_frequencies} entails two different scenarios depending on the spacetime dimension. If $n$ is odd
\begin{equation}
\omega_{k,0,+}\in\mathbb{Z}\Longrightarrow M^2=p^2-\frac{(n-1)^2}{4},
\end{equation}
where $p$ is any integer. If $n$ is instead even
\begin{equation}
\omega_{k,0,+}\in\mathbb{Z}\Longrightarrow M^2=\frac{1}{4}((2p+1)^2+(n-1)^2),
\end{equation}
where $p$ is still integer valued. Observe that, for $n=4$ we reproduce the result in \cite{Avis:1977yn}.
\item Imposing Neumann boundary condition, \eqref{eq:N_frequencies} entails two different scenarios depending on the spacetime dimension. Taking into account the constraint $0<\nu<1$ where $\nu$ is defined in \eqref{eq:nu}, then, if $n$ is odd there exists no admissible mass. On the contrary, if $n$ is even, there is only one admissible possibility:
\begin{equation}
\nu=\frac{1}{2}\Longrightarrow M^2=-\frac{n^2-2n}{4}
\end{equation}
\item Imposing an arbitrary Robin boundary condition, that is choosing $\gamma\in(0,\pi)$ with $\gamma\neq\frac{\pi}{2}$, one can realize from \eqref{eq:tandelta} that, if $\gamma\neq 0,\frac{\pi}{2}$, there exists no value of $\nu\in(0,1)$ for which the function is periodic for integer values of $2\pi\omega$, regardless of the dimension $n$. This can be realized by assuming that, for a given boundary condition, the solutions are periodic with integer period and exploiting that the Euler Gamma functions enjoy the recursion relation $\Gamma(z+1)=z\Gamma(z)$. Hence $\lambda_{2,\gamma}$ does not induce in these cases an admissible counterpart in AdS$_n$.
\end{enumerate}


\section{Conclusions}

In this paper we have discussed the class of boundary conditions which can be assigned to a massive, real scalar field on the global patch of anti-de Sitter spacetime. Working with the universal cover CAdS$_n$, we have shown that one can consider the full family of Robin boundary conditions and, to each of them, one can assign an explicit two-point correlation function which enjoys the Hadamard property. In addition we have proven that, unless one considers the Dirichlet case (or in one instance also the Neumann one), none of these two-point functions admits a well-behaved counterpart on AdS$_n$. 

This work supports the relevance of studying under full-generality the possible class of boundary conditions which can be associated to a field theory when dealing with manifolds with a boundary. In this respect it would be interesting to consider on CAdS$_n$ more general scenarios, such as dynamical boundary conditions which have been recently studied in the Poincar\'e patch in \cite{Dappiaggi:2018pju} and in \cite{DDF} from a rigorous viewpoint.


\appendix

\section{Eigenfunction representation of the delta-distribution}

Goal of this Appendix is to construct $\widehat{\lambda}_{2,L,\omega}$ starting from \eqref{eq:delta_form} and from $S$ the operator \eqref{eq:SL} which represents the Sturm-Liouville form of \eqref{eq:KG}. A convenient and equivalent way of addressing this question consists of recasting \eqref{eq:SL} as  $S_0\Phi_\omega=\omega^2\mathcal{J}(z)\Phi_\omega$ where 
$$S_0=\frac{\dd}{\dd z}\left(P(z)\frac{\dd}{\dd z}\right)+\widetilde{Q}(z).$$
This can be read as an eigenvalue problem for the symmetric operator $S_0$ on the Hilbert space $L^2((0,1),\dd\mu(z))$ where $\dd\mu(z)=\mathcal{J}(z)\dd z$ and where $\omega^2$ plays the role of the spectral parameter. Hence, in this setting our original problem boils down to finding a resolution of the identity operator in terms of eigenfunctions of $S_0$. Most notably, as first discussed in \cite{Ishibashi:2004wx} and then applied in \cite{Dappiaggi:2016fwc}, there exists one such resolution for each self-adjoint extension of the operator $S_0$. Recollecting the results of \cite{Ishibashi:2004wx}, it turns out that $S_0$ is essentially self-adjoint if $\nu\geq 1$, $\nu$ being defined in \eqref{eq:nu}. In this case there exists only one self-adjoint extension and an associated unique resolution of the identity. In the language of differential equations this amounts to saying that no boundary condition should be imposed when solving \eqref{eq:ODE}. On the contrary, if $0<\nu<1$, there exists a one-parameter family of self-adjoint extensions of $S_0$ which can be parametrized in terms of a boundary condition at $z=1$ of the form \eqref{eq:boundary_condition}. Hence, for each $\gamma\in[0,\pi)$, there exists a different resolution of the identity. 

\begin{figure*}[t!]
	\centering
	\includegraphics[width=0.415\linewidth]{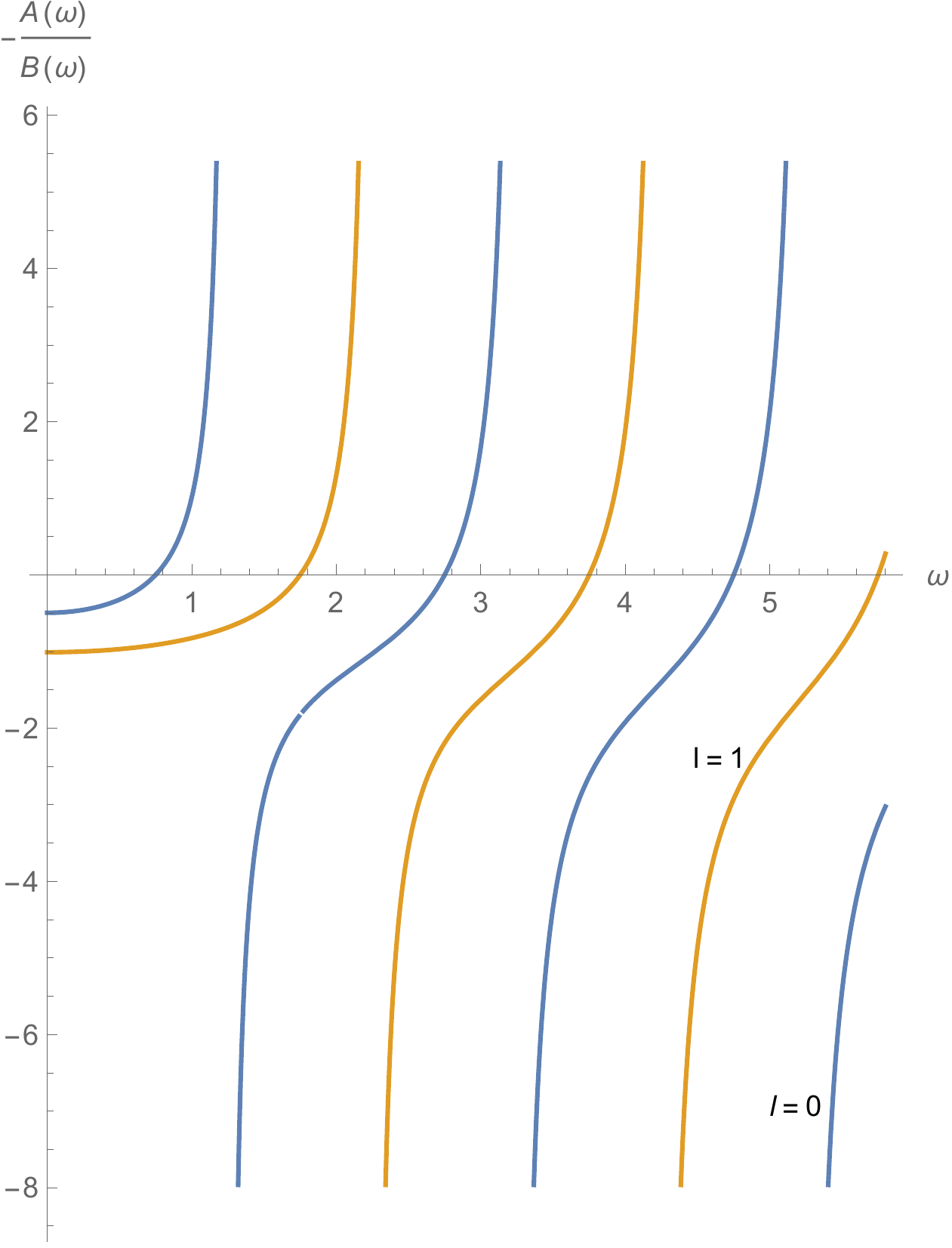} \hspace*{5ex}
	\includegraphics[width=0.415\linewidth]{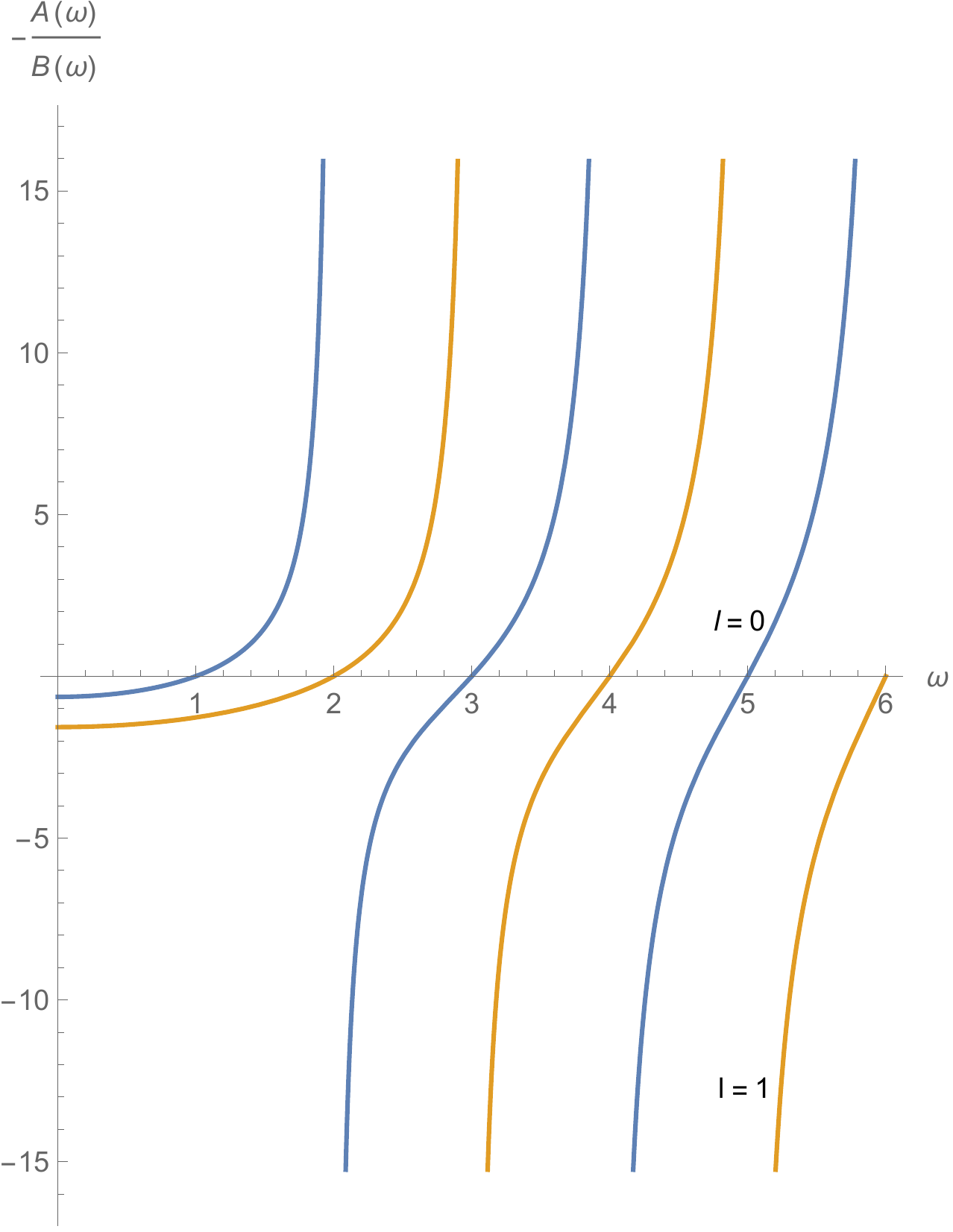}
	\caption{\label{plot_solutions}Plot of \eqref{eq:tandelta} for $n=3$, $\nu=\frac{1}{2}$ (left figure) and $n=3$, $\nu=\frac{1}{4}$ (right figure). In both cases we consider $l_3=l_4=0,1$ and only positive frequencies since \eqref{eq:tandelta} enjoys the symmetry $\omega\to-\omega$.}
\end{figure*}

The translation of the above reasoning into an explicit construction is well-understood, \cite[Chap. 7]{Stakgold}. The first step consists of constructing the Green's operator associated to \eqref{eq:SL}. Hence, focusing on the case $\nu\in(0,1)$, for each $\gamma\in[0,\pi)$, we look for a bi-distribution $\mathcal{G}_{S_0,\omega,\gamma}$, $\gamma\in[0,\pi)$ whose integral kernel obeys to
\begin{widetext}
\begin{eqnarray}\label{eq:Green_def}
\left((S_0-\omega^2\mathbb{I})\otimes\mathbb{I}\right)\mathcal{G}_{S_0,\omega,\gamma}(z,z^\prime)=\left(\mathbb{I}\otimes(S_0-\omega^2\mathbb{I})\right)\mathcal{G}_{S_0,\omega,\gamma}(z,z^\prime)=\frac{\delta(z-z^\prime)}{\mathcal{J}(z)}.
\end{eqnarray}
Since $S$ is an ordinary differential operator, standard techniques yield
\begin{equation}\label{eq:Green_ODE}
\mathcal{G}_{S_0,\omega,\gamma}(z,z^\prime)=\mathcal{N}_\omega\left(\Theta(z-z^\prime)\Phi_{1,\omega}(z)\Phi_{\gamma,\omega}(z^\prime)+\Theta(z^\prime-z)\Phi_{\gamma,\omega}(z)\Phi_{1,\omega}(z^\prime)\right),
\end{equation}
\end{widetext}
where $\Phi_{1,\omega}$ is the solution \eqref{Phi1}, $\Phi_{\gamma,\omega}$ that in \eqref{Phi_gamma}, while $\Theta$ is the Heaviside distribution. The remaining normalization constant can be computed directly from \eqref{eq:Green_def} using the connection formulae for Kummer's solutions \cite[15.10.17 \& 15.10.18]{NIST} being
\begin{gather}
\mathcal{N}^{-1}_\omega=P(z)W_z[\Phi_1(z),\Phi_\gamma(z)]=\notag\\
=-2(\cos\gamma A(\omega)+\sin\gamma B(\omega)),\label{eq:normalization}
\end{gather}
where $P(z)=-Q_1(z)\mathcal{J}(z)$, {\it cf.} \eqref{eq:SL}, while
\begin{subequations}
	\begin{align}
	A(\omega) = \frac{\Gamma(c)\Gamma(a+b-c+1)}{\Gamma(a)\Gamma(b)}\label{eq:A}\\
	B(\omega) = \frac{\Gamma(c)\Gamma(c-a-b+1)}{\Gamma(c-a)\Gamma(c-b)},\label{eq:B}
	\end{align}
\end{subequations}
where $a,b,c$ are defined in \eqref{eq:a}, \eqref{eq:b} and \eqref{eq:c} respectively. In these formulae we decided for later convenience to make explicit the dependence of $A$ and $B$ on $\omega$ through the coefficients $a$ and $b$. Starting from \eqref{eq:Green_ODE}, the following identity holds true
\begin{equation}\label{eq:contour_integral}
\frac{\delta(z-z^\prime)}{\mathcal{J}(z)}=\frac{i}{2\pi}\oint_{C_{\omega^2}}d(\omega^2)\, \mathcal{G}_{S,\omega,\gamma}(z,z^\prime),
\end{equation}
where $\oint_{C_{\omega^2}}$ indicates that we are considering a contour integral in the complex plane with respect to the spectral parameter $\omega^2$, {\it cf.} \cite{Dappiaggi:2016fwc} and \cite{Stakgold}. A direct inspection of \eqref{eq:Green_ODE} and of \eqref{eq:normalization} unveils that this integral can be solved using Cauchy residue theorem. For all admissible values of $\gamma$, the integrand contains a countable number of simple poles, obtained as the zeros of \eqref{eq:normalization} in terms of $\omega$. It is convenient to distinguish three sub-cases:
\begin{enumerate}
	\item If $\gamma=0$, then $\mathcal{N}^{-1}_\omega=0$ if and only if either $\frac{1}{\Gamma(a)}$ or $\frac{1}{\Gamma(b)}$ vanish. This occurs for a countable set of frequencies, that is ($k\in\mathbb{N}\cup\{0\}$)
	\begin{equation}\label{eq:D_frequencies}
		\omega_{k,0,\pm}=\pm\left(\frac{n-1}{2} + 2 k + l_{n-2} + \nu\right)
	\end{equation}
\item If $\gamma=\frac{\pi}{2}$ then $\mathcal{N}^{-1}_\omega=0$ if and only if either $\frac{1}{\Gamma(c-a)}$ or $\frac{1}{\Gamma(c-b)}$ vanish. This occurs for a countable set of frequencies, that is ($k\in\mathbb{N}\cup\{0\}$)
\begin{equation}\label{eq:N_frequencies}
\omega_{k,\frac{\pi}{2},\pm}=\pm\left(\frac{n-1}{2} + 2 k + l_{n-2} - \nu\right)
\end{equation}
\item if $0<\gamma<\pi$ and $\gamma\neq\frac{\pi}{2}$, then one has to solve in terms of $\omega$ the equation 
\begin{gather}
\cot\gamma=-\frac{B(\omega)}{A(\omega)}=\notag\\
=-\frac{\Gamma(c-a-b+1)\Gamma(a)\Gamma(b)}{\Gamma(a+b-c+1)\Gamma(c-a)\Gamma(c-b)}\label{eq:tandelta}
\end{gather}
Only a numerical evaluation is possible, but one can nonetheless infer that there exists a countable set of such solutions. As a matter of fact the right hand side of \eqref{eq:tandelta}, seen as a function of $\omega$, is continuous, it vanishes whenever $\omega=\omega_{k,0,\pm}$ while it diverges if $\omega=\omega_{k,\frac{\pi}{2},\pm}$. A direct inspection of \eqref{eq:D_frequencies} and of \eqref{eq:N_frequencies} unveils in addition that, for all $k\in\mathbb{N}\cup\{0\}$, 
$$\lim_{\omega\to\omega_{k,\frac{\pi}{2},\pm}^+}\frac{A(\omega)}{B(\omega)}=-\lim_{\omega\to\omega_{k,\frac{\pi}{2},\pm}^-}\frac{A(\omega)}{B(\omega)}.$$
Combining such data together it turns out that \eqref{eq:tandelta} admits a countable number of solutions. In addition, observing that \eqref{eq:A} and \eqref{eq:B} are invariant under the map $\omega\mapsto-\omega$, we can enumerate these solutions as $\omega_{k,\gamma,\pm}$ with $k\in\mathbb{N}\cup\{0\}$ where $\pm$ divides between the positive and the negative ones. An exemplification of the behaviour of $-\frac{A(\omega)}{B(\omega)}$ is given in Figure \ref{plot_solutions}.
\end{enumerate}

Recalling that, whenever the Wronskian between two solutions of an ordinary differential equation vanishes, these are linearly dependent, a direct application of Cauchy residue theorem entails that \eqref{eq:contour_integral} becomes  
\begin{widetext}
	\begin{equation}\label{eq:Delta_resolution}
	\frac{\delta(z-z^\prime)}{\mathcal{J}(z)}=\sum\limits_{k=0}^\infty 2\left(\cos\gamma\, C(\omega_{k,\gamma,+})+\sin\gamma\, D(\omega_{k,\gamma,+})\right)\omega_{k,\gamma,+}\Phi_{1,\omega_{k,\gamma,+}}(z)\Phi_{1,\omega_{k,\gamma,+}}(z^\prime)
	\end{equation}
\end{widetext}
where, recalling that $a,b,c$ are defined in \eqref{eq:a}, \eqref{eq:b} and \eqref{eq:c} respectively, it holds \cite[15.10.17 \& 15.10.18]{NIST}
\begin{subequations}
	\begin{align}
		C(\omega)=\frac{\Gamma(c)\Gamma(a+b-c+1)}{\Gamma(a-c+1)\Gamma(b-c+1)}\label{eq:C}\\
		D(\omega)=\frac{\Gamma(c)\Gamma(c-a-b+1)}{\Gamma(1-a)\Gamma(1-b)},\label{eq:D}
	\end{align}
\end{subequations}
Observe that, in the special case of Dirichlet boundary conditions and for $n$ odd, $c,-a\in\mathbb{N}\cup\{0\}$, while, for Neumann boundary conditions both $c,a\in\mathbb{N}\cup\{0\}$. In the first case \eqref{eq:C} vanishes, while, in the second \eqref{eq:D} vanishes. To avoid this pathological situation, when $\gamma=0$ and when $\gamma=\frac{\pi}{2}$, we use instead \cite[15.10.21 \& 15.10.22]{NIST} setting
\begin{subequations}
	\begin{align}
	C_0(\omega)=\frac{\Gamma(c-a)\Gamma(c-b)}{\Gamma(c)\Gamma(c-a-b)}\label{eq:C1}\\
	D_\frac{\pi}{2}(\omega)=\frac{\Gamma(a)\Gamma(b)}{\Gamma(c)\Gamma(a+b-c)},\label{eq:D1}
	\end{align}
\end{subequations}

It is instructive, thus, to write explicitly the resolution of the Dirac delta in the two special cases, namely the Dirichlet boundary condition $\gamma=0$
\begin{gather}
\frac{\delta(z-z^\prime)}{\mathcal{J}(z)}=
\sum\limits_{k=0}^\infty 2C_0(\omega_{k,0,+})\omega_{k,0,+}\Phi_{1,\omega_{k,0,+}}(z)\Phi_{1,\omega_{k,0,+}}(z^\prime),\label{eq:Delta_resolution_D}
\end{gather}
where $\omega_{k,0,+}$ are the frequencies in \eqref{eq:D_frequencies} and the Neumann boundary condition
\begin{gather}
\frac{\delta(z-z^\prime)}{\mathcal{J}(z)}=\\
\sum\limits_{k=0}^\infty 2D_{\frac{\pi}{2}}(\omega_{k,\frac{\pi}{2},+})\omega_{k,\frac{\pi}{2},+}\Phi_{1,\omega_{k,\frac{\pi}{2},+}}(z)\Phi_{1,\omega_{k,\frac{\pi}{2},+}}(z^\prime),\label{eq:Delta_resolution_N}
\end{gather}
where $\omega_{k,\frac{\pi}{2},+}$ are the frequencies in \eqref{eq:N_frequencies}. Observe that, in these two cases, we used the symmetry of the hypergeometric function under exchange of its two first arguments. As a last comment we observe that, if we consider a range of masses such that $\nu\geq 1$, $\nu$ being defined in \eqref{eq:nu} then the same procedure employed above would yield only one possible resolution of the Dirac delta distribution, namely \eqref{eq:Delta_resolution_D}.

\section*{Acknowledgments}
The authors are grateful to Igor Khavkine and to Jorma Louko for the interesting discussions. This work is based partly on the MSc.~thesis in Physics of A.~M. at the University of Milan. The work of C.~D.\ was supported by the University of Pavia. The work of H.~F.\ was supported by the INFN postdoctoral fellowship ``Geometrical Methods in Quantum Field Theories and Applications'', and in part by a fellowship of the ``Progetto Giovani GNFM 2017 -- Wave propagation on lorentzian manifolds with boundaries and applications to algebraic QFT'' fostered by the National Group of Mathematical Physics (GNFM-INdAM).



\end{document}